# Gravitational Radiation from Oscillating Gravitational Dipole

Fran De Aquino

Maranhao State University, Physics Department, S.Luis/MA, Brazil.
deaquino@uema.br

**Abstract**. The concept of Gravitational Dipole is introduced starting from the recent discovery of *negative* gravitational mass (gr-qc/0005107 and physics/0205089). A simple experiment, a gravitational wave transmitter, to test this new concept of gravitational radiation source is presented.

## 1. INTRODUCTION

When the *gravitational field* of an object *changes*, the changes ripple outwards through space and take a finite time to reach other objects. These ripples are called *gravitational radiation* or *gravitational waves*.

The existence of gravitational waves follows from the General Theory of Relativity. In Einstein's theory of gravity the gravitational waves propagate at the speed of light.

Just as electromagnetic waves (EM), gravitational waves (GW) too carry energy and momentum from their sources. Unlike EM waves, however, there is no dipole radiation in Einstein's theory of gravity. The dominant channel of emission is quadrupolar. But the recent experimental discovery of *negative* gravitational mass[1,2] suggest the possibility of dipole radiation.

This fact is highly relevant because a gravitational wave transmitter can be designed to generate detectable levels of gravitational radiation in the laboratory.

Here, we will study the theory and design of the oscillating gravitational dipole.

## 2. THEORY

Gravity is related to *gravitational mass* of the particles. The physical property of mass has two distinct aspects, *gravitational mass* $m_g$ and *inertial mass* $m_i$. The inertial mass is the mass factor in Newton's 2nd Law of Motion $(F = m_i a)$, while the gravitational mass produces and responds to gravitational fields. It supplies the mass factors in Newton's famous inverse-square law of gravity ($F_{12} = G m_{g1} m_{g2} / r_{12}^2$; $G$ is the Newton's gravitational constant).

According to the *weak* form of Einstein's General Relativity *equivalence principle*, the gravitational and inertial masses are equivalent. However recent calculations [3] have revealed that they are correlated by an adimensional factor, which depends on the incident radiation upon the particle. It was shown that there is a direct correlation between the radiation absorbed by the particle and its gravitational mass, independently of the inertial mass. It was also shown that only in the absence of electromagnetic radiation this factor becomes equal to *one* and that, in specific electromagnetic conditions, it can be reduced, nullified or made negative. This means that we can reduce, nullify or make *negative* the gravitational mass of a body. This unexpected theoretical result has been confirmed by two experiments using Extremely Low Frequency (ELF) radiation upon ferromagnetic material[1,2].

The fact that *negative gravitational mass* to have been detected in both experiments suggest that we now have available a plausible procedure for building an oscillating gravitational dipole.

The general equation of correlation between gravitational and inertial mass can be written as [2]:

$$m_g = m_i - 2\left\{\sqrt{1+\left\{\frac{aD}{m_i c}\sqrt{\frac{\mu\sigma}{4\pi f^3}}\right\}^2} - 1\right\}m_i \quad [1]$$

where $D$ is the *power density* of the incident( or emitted) radiation; $f$ is the frequency of the radiation; $a$ is the area of the surface of the particle of mass $m_i$; $\mu$ and $\sigma$ are respectively, the *permeability* and the *conductivity* of the medium around the particle, in which the incident radiation is propagating. For an *atom* inside a body, the incident(or emitted) radiation upon the atom will be propagating inside the body, and consequently, $\sigma = \sigma_{body}$, $\mu = \mu_{body}$.

Equation(1) shows that, elementary particles (mainly *electrons*) can have their *gravitational masses* strongly reduced by means of Extremely-Low Frequency (ELF) radiation.

Let us consider an electric current $I$ through a conductor (*annealed iron wire* 99.98%Fe; $\mu = 350,000\,\mu_0; \sigma = 1.03 \times 10^7\,S/m$) submitted to ELF electromagnetic radiation with power density $D$ and frequency $f$.

Under these circumstances the *gravitational mass* $m_{ge}$ of the *free electrons* (electric current), according to Eq.(1), is given by

$$m_{ge} = m_e - 2\left\{\sqrt{1+\left\{\frac{a_e D}{m_e c}\sqrt{\frac{\mu\sigma}{4\pi f^3}}\right\}^2} - 1\right\}m_e \quad [2]$$

Here, $\mu$ and $\sigma$ are respectively, the *permeability* and the *conductivity* of the *annealed iron wire*; $m_e = 9.11\times 10^{-31}\,kg$.

If the ELF electromagnetic radiation come from a half-wave *electric* dipole encapsulated by an *annealed iron* (purified iron, with the same characteristics of the *annealed iron wire*), the *radiation resistance* of the antenna for a frequency $\omega = 2\pi f$, can be written as follows [4]

$$R_r = \frac{\omega\mu\beta}{6\pi}\Delta z^2 \quad [3]$$

where $\Delta z$ is the length of the dipole and

$$\beta = \omega\sqrt{\frac{\varepsilon\mu}{2}\left(\sqrt{1+(\sigma/\omega\varepsilon)^2}+1\right)} =$$
$$= \frac{\omega}{c}\sqrt{\frac{\varepsilon_r\mu_r}{2}\left(\sqrt{1+(\sigma/\omega\varepsilon)^2}+1\right)} =$$
$$= \frac{\omega}{c}(n_r) = \frac{\omega}{c}\left(\frac{c}{v}\right) = \frac{\omega}{v} \quad [4]$$

where

$$v = \frac{c}{\sqrt{\frac{\varepsilon_r\mu_r}{2}\left(\sqrt{1+(\sigma/\omega\varepsilon)^2}+1\right)}} \quad [5]$$

is the velocity of the electromagnetic waves through the iron. $(\mu_r = \mu/\mu_0; \varepsilon_r = \varepsilon/\varepsilon_0)$.

Substituting (4) into (3) gives

$$R_r = \frac{2\pi}{3}\left(\frac{\mu}{v}\right)(\Delta z f)^2 \quad [6]$$

Note that when the medium surrounding the dipole is *air* and $\omega \gg \sigma/\varepsilon$, $\beta \cong \omega\sqrt{\varepsilon_0\mu_0}$, $v \cong c$ and $R_r$ reduces to the well-know expression $R_r \cong (\Delta z\omega)^2/6\pi\varepsilon_0 c^3$.

For $\sigma \gg \omega\varepsilon$ the Eq.(6) can be rewritten in the following form

$$R_r = (\Delta z)^2\sqrt{\left(\frac{\pi}{9}\right)\sigma\mu^3 f^3} \quad [7]$$

The *ohmic resistance* of the dipole is [5]

$$R_{ohmic} \cong \frac{\Delta z}{2\pi r_0}R_s \quad [8]$$

where $r_0$ is the radius of the cross

section of the dipole, and $R_S$ is the *surface resistance*,

$$R_S = \sqrt{\frac{\omega \mu_{dipole}}{2\sigma_{dipole}}} \qquad [9]$$

Thus,

$$R_{ohmic} \cong \frac{\Delta z}{r_0}\sqrt{\frac{\mu_{dipole} f}{4\pi\sigma_{dipole}}} \qquad [10]$$

Where $\mu_{dipole} = \mu_{copper} \cong \mu_0$ and $\sigma_{dipole} = \sigma_{copper} = 5.8 \times 10^7 S/m$.

The *radiated power* for an *effective* (*rms*) current $I$ is then $P = R_r I^2$ and consequently, the *power density*, $D$, of the emitted ELF radiation, is

$$D = \frac{P}{S} = \frac{(\Delta z I)^2}{S}\sqrt{\left(\frac{\pi}{9}\right)\sigma\mu^3 f^3} \qquad [11]$$

where $S$ is the area surround of the dipole.

For $f = 69.4 \mu Hz$, the length of the dipole is

$$\Delta z = \lambda/2 = v/2f = \sqrt{\pi/\mu\sigma f} = 0.10 m$$

Substitution of (11) into (2) yields

$$m_{ge} = m_e - 2\left\{\sqrt{1 + \left(\mu^4\sigma^2\left(\frac{a_e}{6 m_e c S}\right)^2 (\Delta z I)^4\right)} - 1\right\} m_e \qquad [12]$$

Note that the equation above doesn't depends on $f$.

Thus, assuming that the *radius* of the electron[6] (region which electron is "concentrated") is $r_e = (1/4\pi\varepsilon_0)(e^2/m_e c^2) = 2.8\times 10^{-15} m$,

$a_e = 4\pi r_e^2 = 9.8\times 10^{-29} m^2$ then the Eq.(12) becomes

$$m_{ge} = m_e - 2\left\{\sqrt{1 + \frac{1.41\times 10^{-6}}{S^2}I^4} - 1\right\} m_e \qquad [13]$$

Thus, for $S \cong 0.1 m^2$ and $I \cong 100 A$ ( current trough the ELF antenna) the gravitational mass of the free-electrons becomes

$$m_{ge} \cong -234.5 m_e$$

This means that they becomes "*heavy*" electrons.

Gravitational effects produced by ELF electromagnetic radiation upon the electric current in a conductor was recently studied [7]. An apparatus has been constructed to test the behavior of current subjected to ELF radiation. The experimental results show that *gravitational* mass of the free-electrons can becomes *strongly negative*.

The oscillation of the gravitational masses of the free-electrons through the wire produces gravitational radiation, but too weak due to the gravitational mass of the electrons to be very small. However when the electrons become "*heavy*", the gravitational radiation flux can be very large.

Consider a *half-wave* electric dipole whose elements are two cylinders of *annealed iron* (99.98%Fe; $\mu = 350,000 \mu_0; \sigma = 1.03\times 10^7 S/m$) subjected to ELF radiation with frequency $f = 69.4 \mu Hz$ (see Fig.1). The *energy* flux carried by the emitted gravitational waves can be estimated by analogy to the oscillating *electric* dipole.

As we know, the *intensity* of the emitted *electromagnetic* radiation from an oscillating electric dipole ( i.e., the energy across the area unit by time unit in the direction of propagation) is given by [8]

$$F(\phi) = \frac{\pi^2 \Pi_0^2 f^4}{2c^3\varepsilon r^2} \sin^2\phi \qquad [14]$$

The *electric dipole moment* , $\Pi = \Pi_0 \sin\omega t$, can be written as $qz$, where $q$ is the oscillating electric charge, and $z = z_0 \sin\omega t$; thus, one can substitute $\Pi_0$ by $qz_0$, where $z_0$ is the amplitude of the oscillations of $z$.

There are several ways to obtain the equivalent equation for the *intensity* of the emitted *gravitational* radiation from an *oscillating*

*gravitational dipole*. The simplest way is merely the substitution of $\varepsilon$ (*electric permittivity*) by $\varepsilon_G = 1/16\pi G$ (*gravitoelectric permittivity* [9] ) and $q$ by $m_g$ (by analogy with electrodynamics, the *gravitoelectric dipole moment* can be written as $m_g z$, where $m_g$ is the oscillating gravitational mass). Thus the *intensity* of the emitted *gravitational* radiation from an oscillating *gravitational* dipole, $F_{gw}(\phi)$, can be written as follows:

$$F_{gw}(\phi) = \frac{8\pi^3 G m_g^2 z_0^2 f_{gw}^4}{c^3 r^2} sin^2\phi \qquad [15]$$

where $f_{gw}$ is the frequency of the gravitational radiation (equal to the frequency of the electric current through the dipole).

Similarly to the electric dipole, the intensity of the emitted radiation from a gravitational dipole is maximum at the equatorial plane ($\phi = \pi/2$) and zero at the oscillation direction ($\phi = 0$).

The gravitational mass $m_g$ in Eq.(15) refers to the *total* gravitational mass of the "*heavy*" electrons, given by

$$m_g = (10^{29} free-electrons/m^3) V_{ant} m_{ge}$$

where $V_{ant}$ is the volume of the antenna.

For the microwave antenna in Fig.1, $V_{ant} = 7.4 \times 10^{-8} m^3$ and $m_{ge} \cong -234.5 m_e$. This gives $m_g \cong 10^{-6} kg$.

The amplitude of oscillations of the half-wave *gravitational* dipole is

$$z_0 = \lambda_{gw}/2 = \frac{c}{2f_{gw}}$$

This means that to produce gravitational waves with frequency $f_{gw} = 10 GHz$, the length of the dipole, $z_0$, must be equal to 1.5cm. By substitution of these values and $m_g \cong -10^{-6} kg$ into Eq.(15) we obtain

$$F_{gw}(\phi) \cong 10^{-9} \frac{sin^2\phi}{r^2} \qquad [16]$$

At a distance $r = 1m$ from the dipole the maximum value of $F_{gw}(\phi)$ is

$$F_{gw}(\tfrac{\pi}{2}) \cong 10^{-9} W/m^2$$

For comparison, a gravitational radiation flux from astronomical source with frequency 1Hz and amplitude $h \cong 10^{-22}$ ( the dimensionless amplitude $h$ of the gravitational waves of astronomical origin that could be detected on earth and with a frequency of about 1 kHz is between $10^{-17}$ and $10^{-22}$ ) has [10],

$$F_{gw} = \frac{1}{32\pi} \frac{c^3}{G} \left(\frac{dh}{dt}\right)^2 =$$

$$= 1.6 \times 10^{-5} \left(\frac{f}{100 Hz}\right)^2 \left(\frac{h}{10^{-22}}\right)^2 \cong 10^{-9} W/m^2$$

As concerns detection of the gravitational radiation from dipole, there are many options. A similar gravitational dipole can also absorb energy from an incident gravitational wave. If a gravitational wave is incident on the gravitational dipole(receiver) in Fig.1(b) the masses of the "heavy" electrons will be driven into oscillation. The amplitude of the oscillations will be the same of the emitter, i.e., 1.5cm.

Recently superconductors have been considered as macroscopic quantum gravitational antennas and transducers[11], which can directly convert a beam of gravitational radiation into electromagnetic radiation and vice versa.

In short, now there is a strong evidence that will be possible to generate and detect gravitational radiation in laboratory.

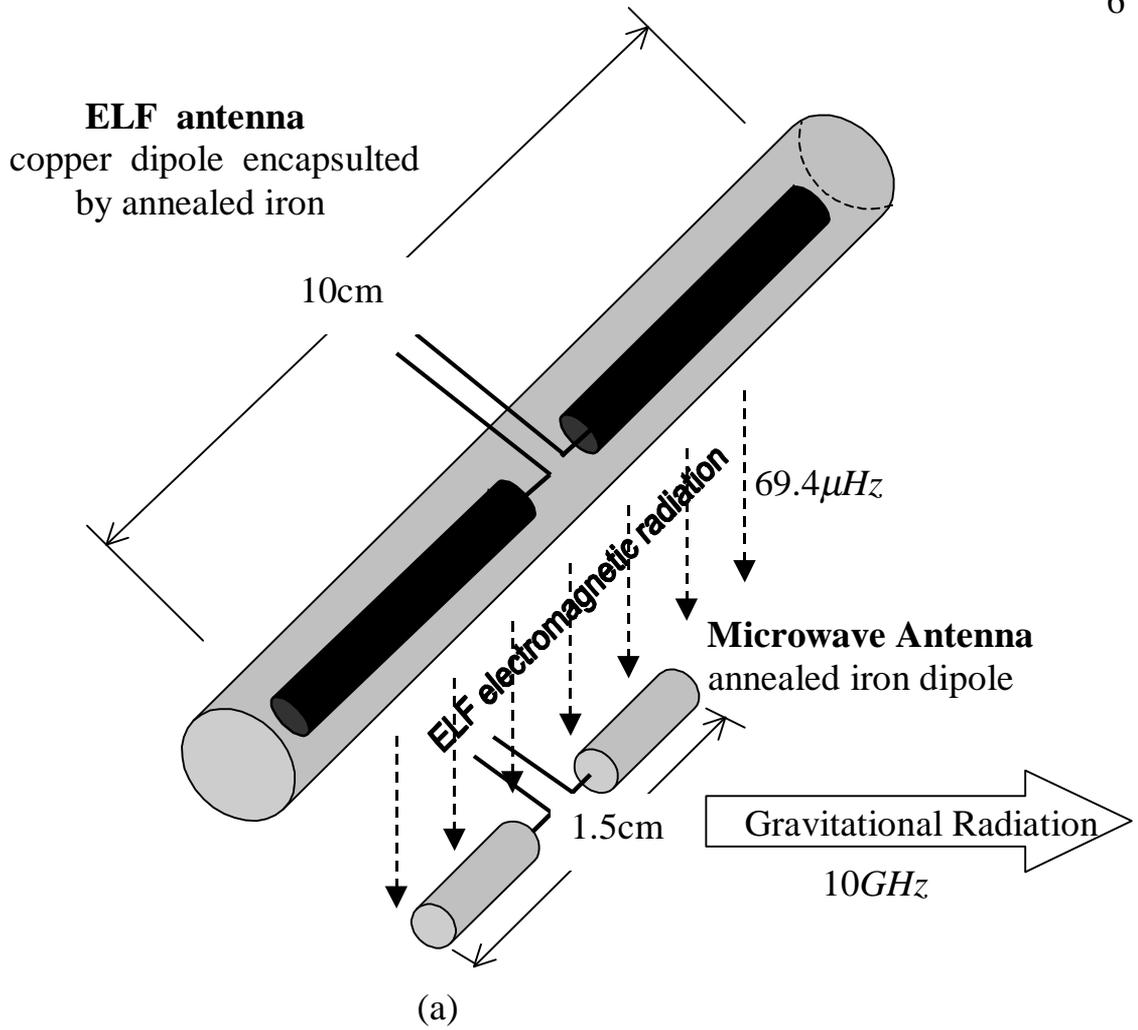

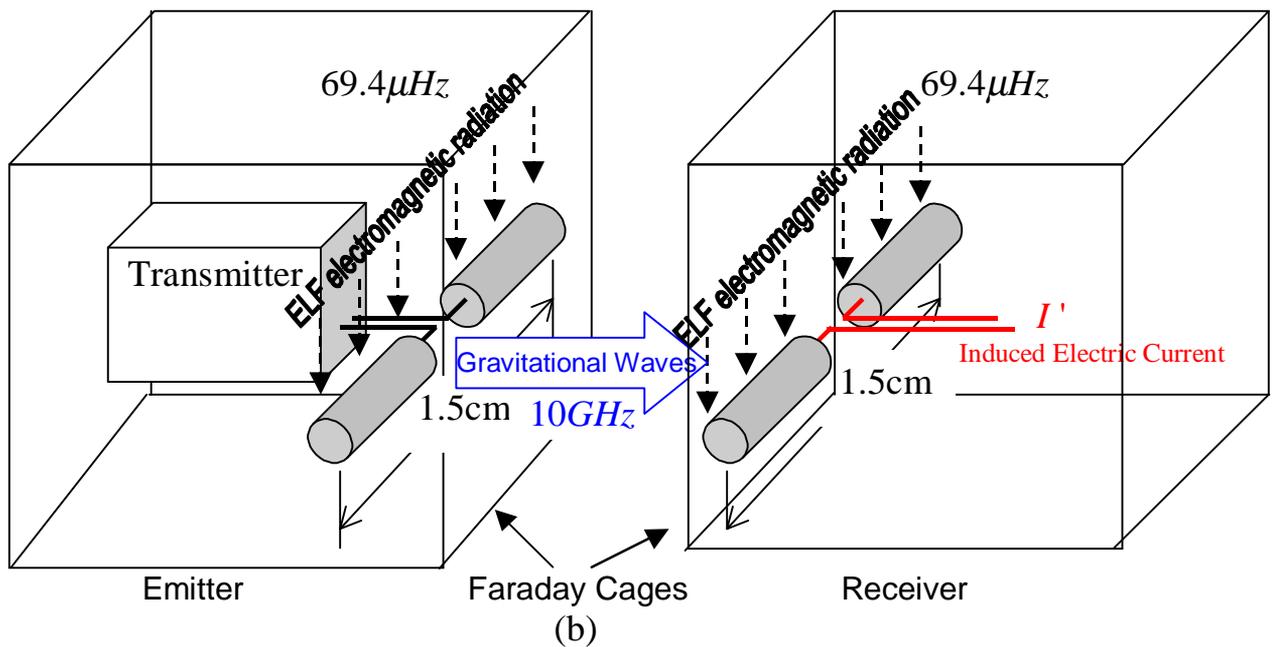

Fig.1 - Schematic diagram of the antennas to produce and receive gravitational radiation.